\documentclass[doublecol]{epl2} 

\usepackage{amsmath}
\usepackage{amsfonts}
\usepackage{braket}
\usepackage{amssymb}

\usepackage{color}
\definecolor{Blue}{rgb}{0.3,0.3,0.9}
\definecolor{Red}{rgb}{0.9,0.3,0.3}

\title{Using entanglement to discern phases in the disordered one-dimensional Bose-Hubbard model}
\shorttitle{Entanglement and the disordered 1D Bose-Hubbard model} 

\author{Andrew M.\ Goldsborough$^{1,2}$ \and Rudolf A.\ R\"{o}mer$^{1}$}
\shortauthor{A.\ M.\ Goldsborough \etal}

\institute{$^{1}$Department of Physics and Centre for Scientific Computing, The University of Warwick, Coventry, CV4 7AL, UK \\
$^{2}$JARA Institute for Quantum Information, RWTH Aachen University, D-52056 Aachen, Germany}


\abstract{
We perform a matrix product state based density matrix renormalisation group analysis of the phases for the disordered one-dimensional Bose-Hubbard model. For particle densities $N/L = 1$, $1/2$ and $2$ we show that it is possible to obtain a full phase diagram using only the entanglement properties, which come \emph{for free} when performing an update. We confirm the presence of Mott insulating, superfluid and Bose glass phases when $N/L = 1$ and $1/2$ (without the Mott insulator) as found in previous studies. For the $N/L = 2$ system we find a double lobed superfluid phase with possible reentrance.} 

\begin{document}

\maketitle

\section{Introduction}
The study of bosons in one dimension has been of great interest in both theoretical and experimental physics for many years due in part to the existence of a quantum phase transition from a superfluid to insulator at zero temperature \cite{GreMEH02}.
The introduction of disorder causes a further phase transition into a localised \emph{Bose glass} phase, which is insulating but remains compressible \cite{FisWGF89}.
The experimental study of phase transitions in bosonic systems is possible using Helium in porous media \cite{CroHST83,CroVR97}, Josephson junction arrays \cite{VanEGM96}, thin films \cite{HavLG89,Pan89} and, more recently, optical lattices \cite{GreMEH02,BakGPF09}.
It is now possible to introduce disorder in a controlled manner to optical lattices using speckle potentials \cite{HorC98,BilJZB08} to study these transitions directly \cite{WhiPMZ09,PasMWD10}.

Analytical results even for clean systems are limited. There is an approximate Bethe-ansatz solution \cite{Kra91}, where the maximum number of bosons per site is set to two. 
For disordered systems Giamarchi and Schulz used renormalization group (RG) techniques to determine the weak disorder physics given the Luttinger parameter $K$ \cite{GiaS87,RisPLG12}.
There are further real-space RG results for the case of strong disorder \cite{AltKPR10,PieA13}.
Numerical approaches provide some of the most effective means of garnering information. 
Quantum Monte Carlo has been employed in 1, 2 and 3 dimensions \cite{ScaBZ91,ProS98,KraTC91,SoyKPS11,GurPPS09,Pol13}, but these methods become difficult in the limit of zero temperature. 
An ideal method for analysing one dimensional systems is the density-matrix RG (DMRG) \cite{Whi92}. It has been applied with great success to a number of physical systems from quantum chemistry \cite{ShaC12} to quantum information\cite{PerVWC07}, including the disordered Bose-Hubbard model \cite{PaiPKR96,RapSZ99}.
The phase diagrams obtained using these methods for the one dimensional case \cite{ProS98,RapSZ99}, whilst qualitatively agreeing, are quantitatively quite different.
This difference could be down to the choice of different observables and the difficulties each has with finite size effects.

In recent years the use of entanglement properties as a means of deciphering phase has become commonplace \cite{LiH08,EisCP10,PolTBO10,DenS11,DenCOM13,KjaBP14}. 
Entanglement is a measurement of a wavefunction's non-locality and as such it is an ideal means of analysing various phases.
Modern numerical techniques such as \emph{tensor networks} and DMRG obtain entanglement information as part of the update algorithms, so large amounts of information about the phase is gathered automatically in the course of the RG iterations \cite{Sch11}.
In this paper we perform a DMRG simulation of the disordered Bose-Hubbard model in the form of a variational update of a matrix product state (MPS) \cite{OstR95,Sch11} implemented in the {\sc ITensor} libraries \cite{Itensor023}.
The disordered Bose-Hubbard model is made up of bosonic creation, $b_{i}^{\dagger}$, and annihilation operators, $b_{i}$, on sites of a linear lattice. 
The Hamiltonian is \cite{RapSZ99}
\begin{equation}
H= - \sum_{i}^{L-1} \frac{t}{2} (b^{\dagger}_{i} b_{i+1} + \mbox{h.c.} ) + \sum_{i}^{L} \frac{U}{2} n_{i}(n_{i}-1) + \mu_{i} n_{i},
\end{equation}
where $n_{i}=b^{\dagger}_{i}b_{i}$ is the local occupation or \emph{number operator} that gives the number of bosons on site $i$. 
The potential disorder is modelled via uniformly distributed random chemical potentials $\mu_i\in [-\Delta \mu/2, \Delta \mu/2]$. 
For ease of comparison, we have adopted prior conventions \cite{RapSZ99} for hopping $t$ and interaction $U$ and throughout the rest of the analysis $t=1$.


\section{Observables}


The Mott insulator can be differentiated from the Bose glass phase by the existence of the \emph{Mott gap}, $E_{g}$, between the ground and first excited 
state. While DMRG ordinarily finds the ground state of the system, low lying excited states have to be constructed iteratively by 
orthogonalising with respect to the lower lying states \cite{Sch11}. For the Bose-Hubbard chain it is numerically more convenient to use the fact that the energy of 
the excited state is equal to the difference in energy between the chemical potential for particle, $\mu_{p} = E_{N+1}-E_{N}$, and hole, $
\mu_{h} = E_{N}-E_{N-1}$, excitations \cite{RapSZ99}.
This means that $E_{g}$ can be found by calculating the energies $E_{N+1}$, $E_{N}$ and $E_{N-1}$ of the $N+1$, $N$ and $N-1$ particle sectors, respectively, as
\begin{equation}
E_{g} = E_{N+1}-2E_{N}+E_{N-1}.
\end{equation}
Hence the determination of $E_{g}$ requires a DMRG run for each of the three different particle numbers and each set of parameters.

The superfluid phase is determined by a non-zero \emph{superfluid fraction} $\rho_{s}$. This is defined as the difference between the ground state energies of a chain with periodic boundaries and anti-periodic boundaries,
\begin{equation}
\rho_{s} = \frac{2L^{2}}{\pi^{2}N} \left( E_{N}^{\mathrm{anti-periodic}} - E_{N}^{\mathrm{periodic}} \right),
\end{equation}
where $L$ is the chain length and $N$ the number of bosons \cite{RapSZ99}. 
For Mott insulator and Bose glass phases, we have $\rho_{s}=0$, so a finite $\rho_{s}$ indicates superfluidity in the phase diagram. From a computational point of view, $\rho_{s}$ is not an easy quantity to determine as it requires the use of periodic boundaries, which are well-known to converge slower and be less accurate than for open systems when using DMRG \cite{Sch11}. Furthermore, as $\rho_{s}$ is the difference between two energies, two such periodic DMRG calculations have to be performed for each set of parameters.

The two-point correlation function $\langle b_{i}^{\dagger} b_{j} \rangle$ 
provides information regarding the localisation of the wavefunction. For the Bose glass and Mott insulating phases the correlation function decays exponentially,  
$\langle \langle b_{i}^{\dagger} b_{j} \rangle \rangle \propto e^{-|i-j|/\xi}$,  
where $\xi$ is the correlation length and $\langle \langle \dots \rangle \rangle$ denotes the expectation value when averaged over all pairs of sites separated by $|i-j|$ and all disorder realisations \cite{RouBMK08}. 
Extended phases like the superfluid are not localised so $\xi$ diverges in the thermodynamic limit.
In the absence of disorder the superfluid phase will be described by Luttinger liquid theory \cite{Voi95}, hence the correlation function will admit a power law decay
\begin{equation}
\langle \langle b_{i}^{\dagger} b_{j} \rangle \rangle \propto |i-j|^{-1/2K},
\label{eq-pld-poly}
\end{equation}
where $K$ is the Luttinger parameter.
$K$ takes the value $2$ for a Kosterlitz-Thouless (KT) transition from superfluid to Mott insulator \cite{Gia97,KuhWM00}.
By utilizing an RG approach, Giamarchi and Schulz \cite{GiaS87} showed that disorder scales to zero in the weak disorder regime when $K>3/2$, giving a superfluid phase. On the other hand,  disorder grows for $K<3/2$ signifying a Bose glass.
This was later extended \cite{RisPLG12} to the \emph{medium disorder} case ($U \sim \Delta\mu$). 
%
Instead of the infinite-system size result \eqref{eq-pld-poly} we use the conformal field theory (CFT) expression \cite{RouBMK08} for an open chain of size $L$,
\begin{equation}
\langle b_{i}^{\dagger} b_{j} \rangle \propto \left[ \frac{\pi}{2L} \frac{ \sqrt{ \left| \sin \left( \frac{\pi i}{L} \right) \right| \left| \sin \left( \frac{ \pi j}{L} \right) \right| }}{  \left| \sin  \frac{ \pi (i+j)}{2L}  \right| \left| \sin  \frac{ \pi (i-j)}{2L}  \right|} \right]^{{1}/{2K}}.
\end{equation}
The expression has to be averaged over all $i$ and $j$ with separation $|i-j|$ and, of course, also averaged over disorder realisations.
Calculating correlation functions does not require multiple DMRG runs, but requires the calculation of an expectation value for each combination of $i$ and $j$, of which there are $L(L-1)/2$.
Furthermore, the accuracy of locating the KT transition from correlation functions for the Bose-Hubbard model has previously been questioned \cite{KuhM98,RouBMK08}.

In each DMRG run, bipartitioning the chain into \emph{system} and \emph{environment} blocks is done routinely to compute singular-value decompositions \cite{Sch11}. 
These singular values, $s_{a}$, can themselves be used to obtain information regarding the phase \cite{PolTBO10,DenS11,DenCOM13} without the need for multiple DMRG runs, thus saving substantial numerical costs. 
The most common such measure is the \emph{entanglement entropy} or \emph{Von Neumann entropy} defined as
\begin{equation}
S_{\mathrm{A}|\mathrm{B}} = -\mathrm{Tr} \rho_{\mathrm{A}} \log_{2} \rho_{\mathrm{A}} = - \sum_{a=1} s_{a}^{2} \log_{2} s_{a}^{2},
\label{eq:entropy}
\end{equation}
which gives the entanglement between regions A and B \cite{Sch11}. 
The reduced density matrix, $\rho_{\mathrm{A}}$, for region A is obtained from the density matrix by tracing over degrees of freedom from region B. Its eigenvalues are given as squares of the $s_a$'s. 
Hence $S_{\mathrm{A}|\mathrm{B}}$ is a measure of the spread of the $s_a$ values. 
If there is one non-zero singular value then the regions are in a product state of the two regions. 
The other extreme is if all singular values are equal, in which case the subsystems are \emph{maximally entangled}. 
In the subsequent analysis we shall average the entanglement entropy over all possible bipartitions along the chain. 
This averaged entanglement entropy can distinguish between phases with high and low entanglement, for example the superfluid and Mott insulating phases.
\begin{figure*}[bt]
\centerline{
(a)\includegraphics[width=0.3\textwidth]{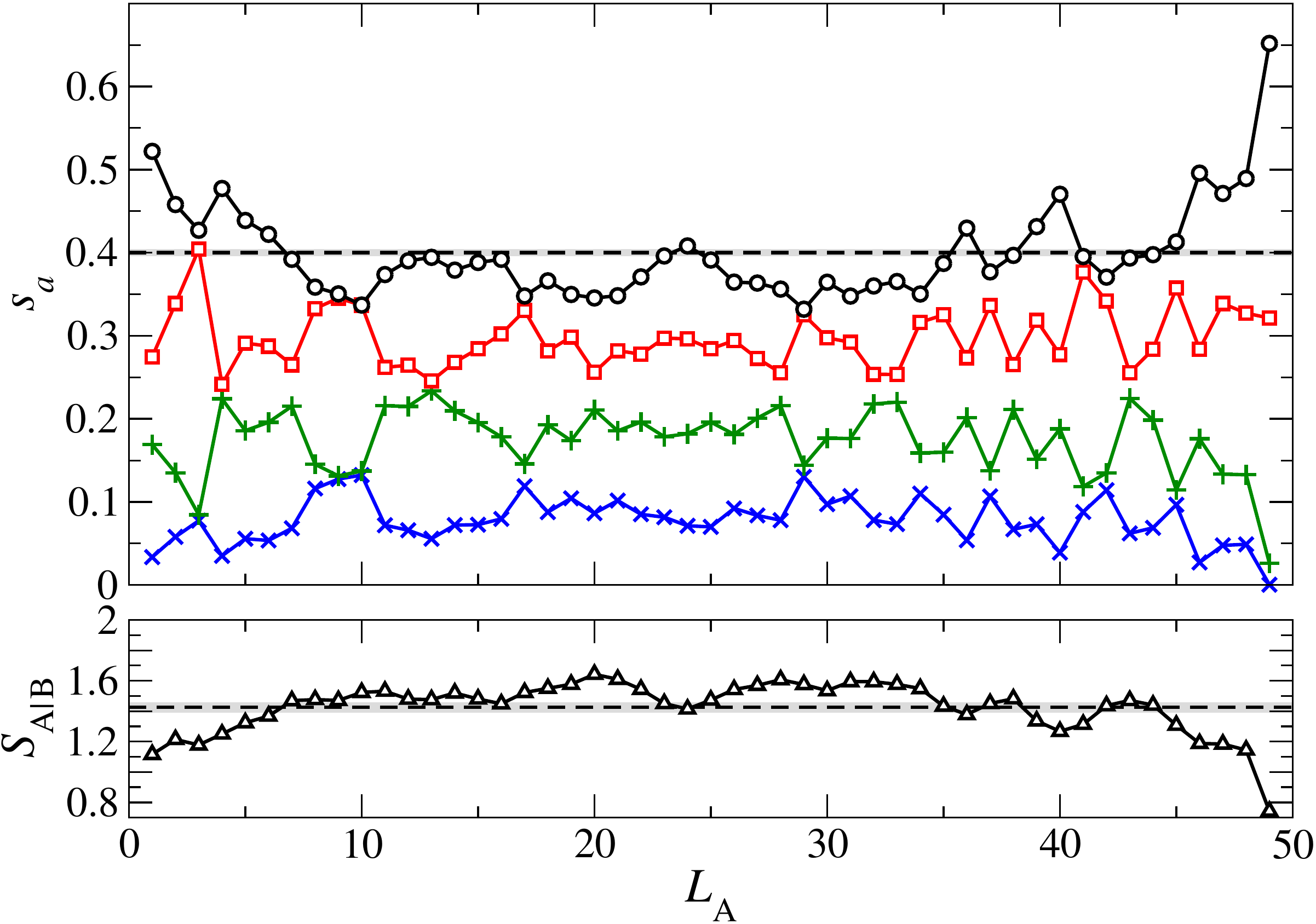}
(b)\includegraphics[width=0.3\textwidth]{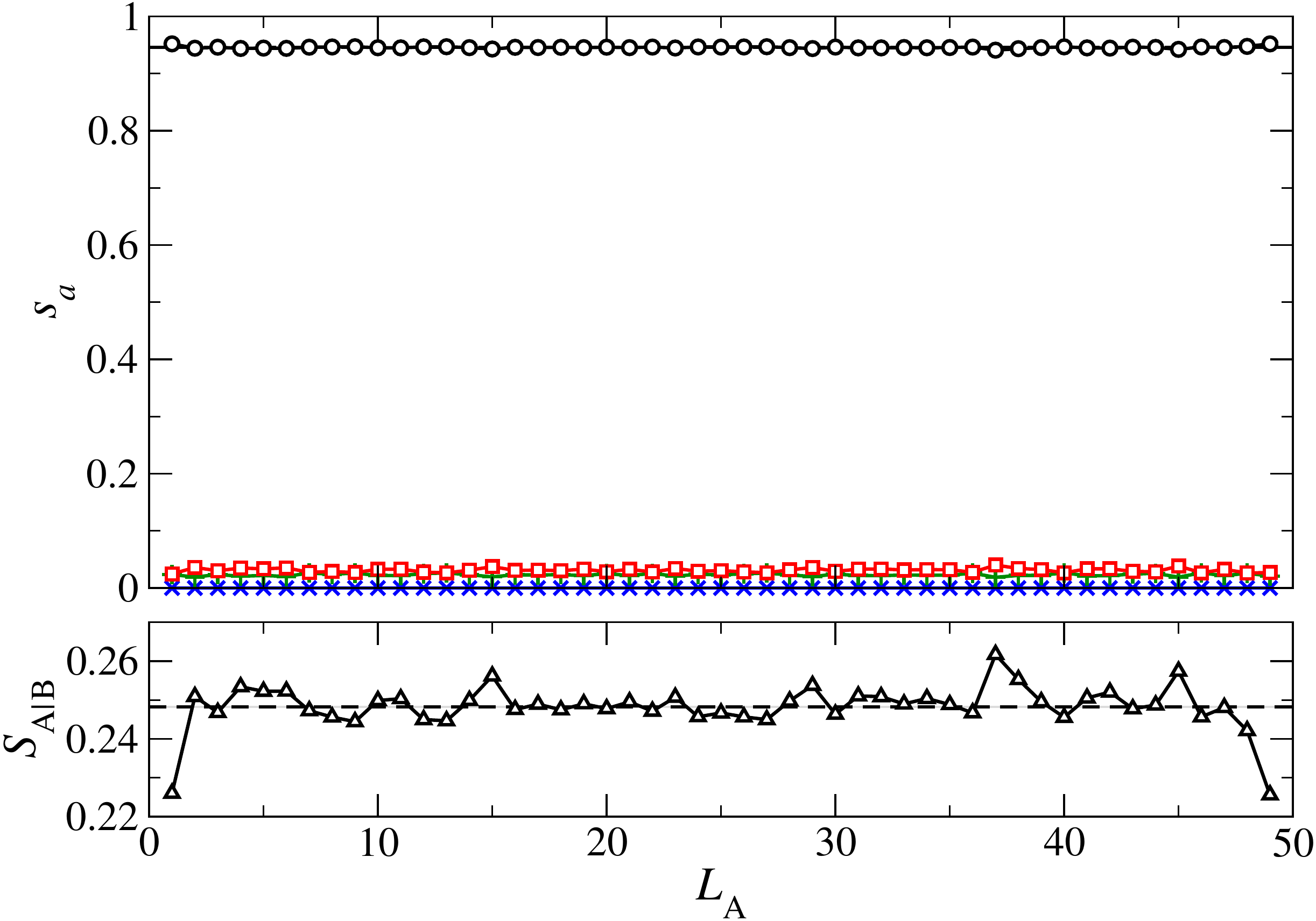}
(c)\includegraphics[width=0.3\textwidth]{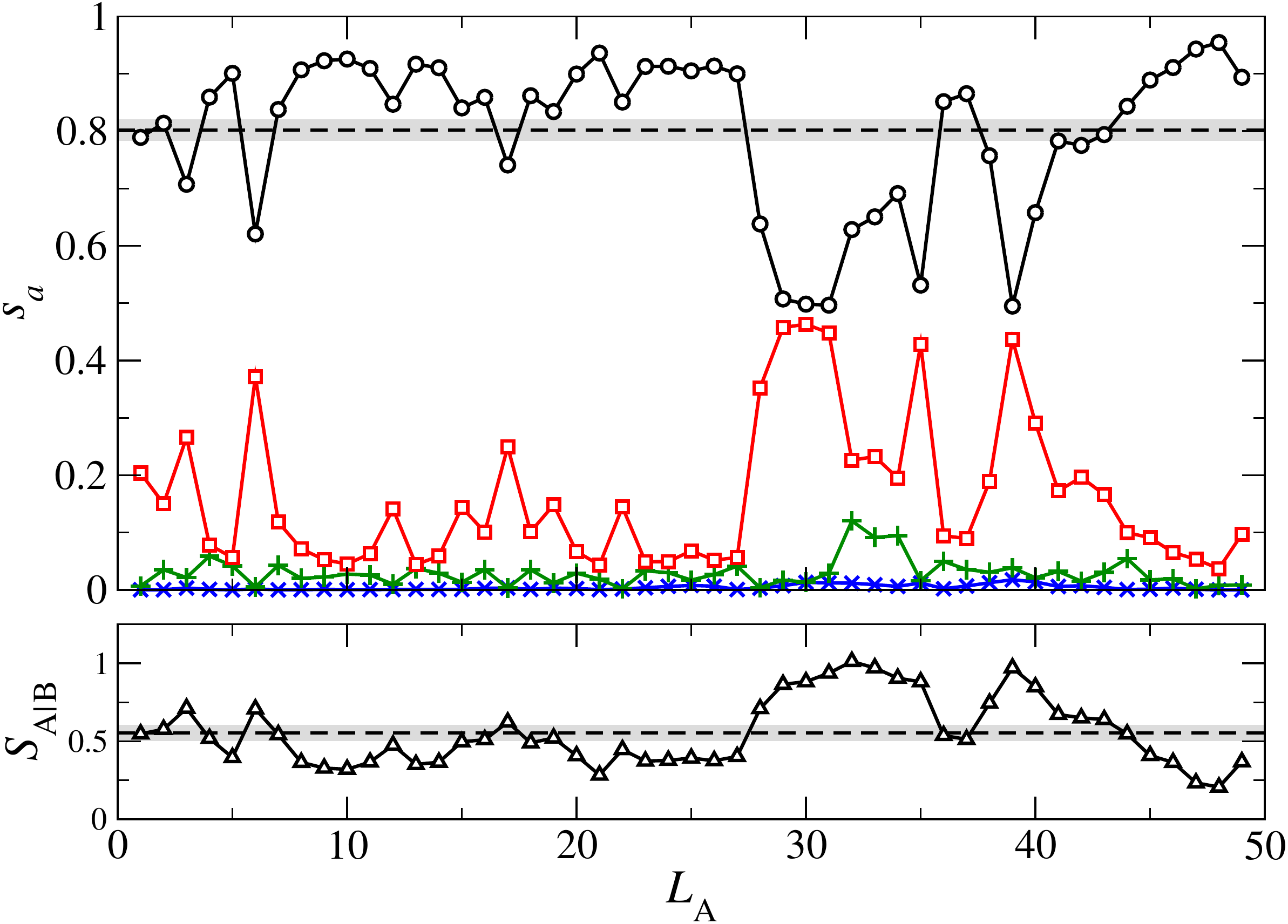}}
\caption{(Color online)
(Top) the largest four singular values, $s_{1}$ (black $\circ$), $s_{2}$ (red $\square$), $s_{3}$ (green $+$), and  $s_{4}$ (blue $\times$) and (Bottom) the entanglement entropy $S_{\mathrm{A}\mathrm{B}}$ (black $\triangle$) for all possible bipartition positions, $L_{\mathrm{A}}$, along a chain of length $L=50$ for
(a) superfluid with $U = 0.5$, $\Delta\mu = 1$, 
(b) Mott insulator with $U = 4.5$, $\Delta\mu = 1$, and
(c) Bose glass with $U = 4.5$, $\Delta\mu = 7$. 
The dashed horizontal line in the top (bottom) graph shows the average value of $s_1$ ($S_{\mathrm{A}|\mathrm{B}}$) while the grey shading indicates its standard deviation when averaged over all $L_{\mathrm{A}}$ positions.
Solid lines connecting symbols are guides to the eye.
}
\label{fig:eespecs}
\end{figure*}

Deng et.\ al.\ \cite{DenCOM13} used the \emph{entanglement spectral parameter}, $\zeta$, to obtain the phase diagram for an extended Bose-Hubbard model. 
The $\zeta$ parameter is defined as the sum of the difference between the first and second, and third and fourth, respectively, eigenvalues $s_a^2$ of $\rho_{\mathrm{A}}$ when averaged over all bipartition positions such that $L_{\mathrm{A}}+L_{\mathrm{B}} = L$, i.e.\
\begin{equation}
\zeta = \overline{\lambda}_{1} - \overline{\lambda}_{2} + \overline{\lambda}_{3} - \overline{\lambda}_{4},
\label{eq:zeta}
\end{equation}
with 
$
\overline{\lambda}_{a} 
\equiv  \sum_{L_{\mathrm{A}}=1}^{L-1} s^{2}_{a}(L_{\mathrm{A}})/({L-1})$,  $a= 1, 2, 3, 4$, 
the bipartition-averaged $a$-th eigenvalue.
In fig.\ \ref{fig:eespecs} we show the typical behaviour of the four lowest $s_{a}$ values in the superfluid, fig.\ \ref{fig:eespecs}(a), Mott insulator, fig.\ \ref{fig:eespecs}(b), and Bose glass, fig.\ \ref{fig:eespecs}(c), regimes
We see that the entanglement spectrum of the superfluid phase is somewhat noisy with the four singular values being of the same order of magnitude. 
Therefore the resulting $S_{\mathrm{A}|\mathrm{B}}$ is large, $\zeta$ is small, but neither have too much variation along $L_{\mathrm{A}}$.
One of the striking features of the entanglement spectrum for the Mott insulator regime is that $s_{1} \approx 1$ while $s_{2} \approx s_{3} \approx s_{4} \approx 0$ for \emph{all} bipartitions, even in the presence of disorder. 
This means that the average $S_{\mathrm{A}|\mathrm{B}}$ will be low and $\zeta \approx 1$, with a negligible deviation. 
Last, entanglement spectra of the Bose-glass show pronounced localised regions separated by areas of low entanglement.
This results in a much larger variation of $\zeta$ and $S_{\mathrm{A}|\mathrm{B}}$ than in the superfluid phase, with $\zeta$ and average $S_{\mathrm{A}|\mathrm{B}}$, between the other phases.
These findings suggest that the average and spatial variations of $S_{\mathrm{A}|\mathrm{B}}$ and $\zeta$ might also be used to distinguish the phases of the disordered Bose-Hubbard model.
%

\section{Results}
\label{results}
We use $S_{\mathrm{A}|\mathrm{B}}$ and $\zeta$ to create qualitative phase diagrams for a modest size of $L=50$, disorder-averaged over $100$ samples using as our DMRG implementation the {\sc ITensor} libraries \cite{Itensor023}. 
The finite-size scaling (FSS) behaviour of $S_{\mathrm{A}|\mathrm{B}}$ and $\zeta$  is currently not well understood, thus to find the phase boundaries for $L\rightarrow\infty$, we perform scaling with $K$ and $E_g$ from estimates up to $L=200$. 
This is a numerically more expensive procedure, so we concentrate on a small number of points with positions motivated by the phase diagram from the entanglement properties.
For disordered systems, getting stuck in local minima is particularly problematic, so we use a relatively large bond dimension $\chi=200$ and perform $20$ DMRG sweeps of the chain for each sample. Our truncation error is less than $10^{-10}$.
We also introduce a small noise term for the first few sweeps; this perturbs a perhaps bad initial wavefunction, allowing faster convergence into the ground state. 

Bosons do not obey the Pauli exclusion principle and hence can condense onto a single site. 
In order to capture such a behaviour, the one-site basis dimension should to be as large as the number of particles in the system. This is numerically infeasible and it is necessary to introduce a finite maximum number of bosons that can occupy each site. 
We use $\max (n_i) = 5$, consistent with ref.\ \cite{KuhWM00} who find that a higher particle number does not effect the results appreciably for $U>0$ \cite{RouBMK08,PaiPKR96,RapSZ99}.\footnote{Five is the current hard limit in the {\sc ITensor} code.} 

\subsection{Density = 1}

For particle density ${N}/{L} = 1$, in the clean case, the system is in a superfluid phase for small $U$ but transitions into a Mott insulating phase at a critical $U_c$. 
Introducing disorder enables the existence of a localized \emph{Bose glass} phase \cite{FisWGF89}. 
The possibility of a direct transition from superfluid to Mott insulator has been discussed extensively (see references in \cite{PolPST09}). In one dimension it was shown \cite{Svi96} that the transition necessarily goes via the Bose glass phase. 
This is now also the accepted picture for any dimension \cite{PolPST09}.

\begin{figure*}[tb]
\center{
(a)\includegraphics[width=0.22\textwidth]{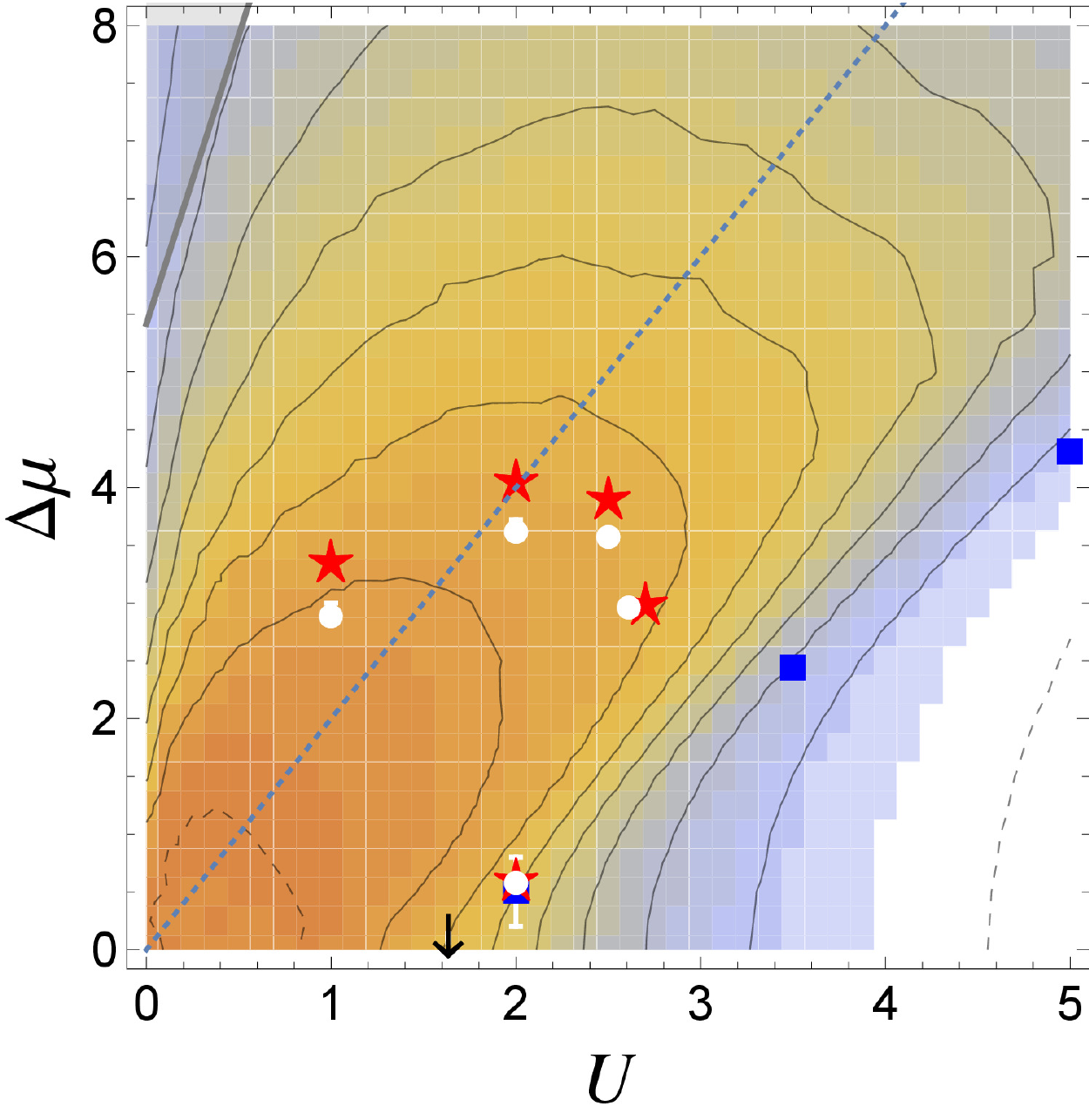}
(b)\includegraphics[width=0.22\textwidth]{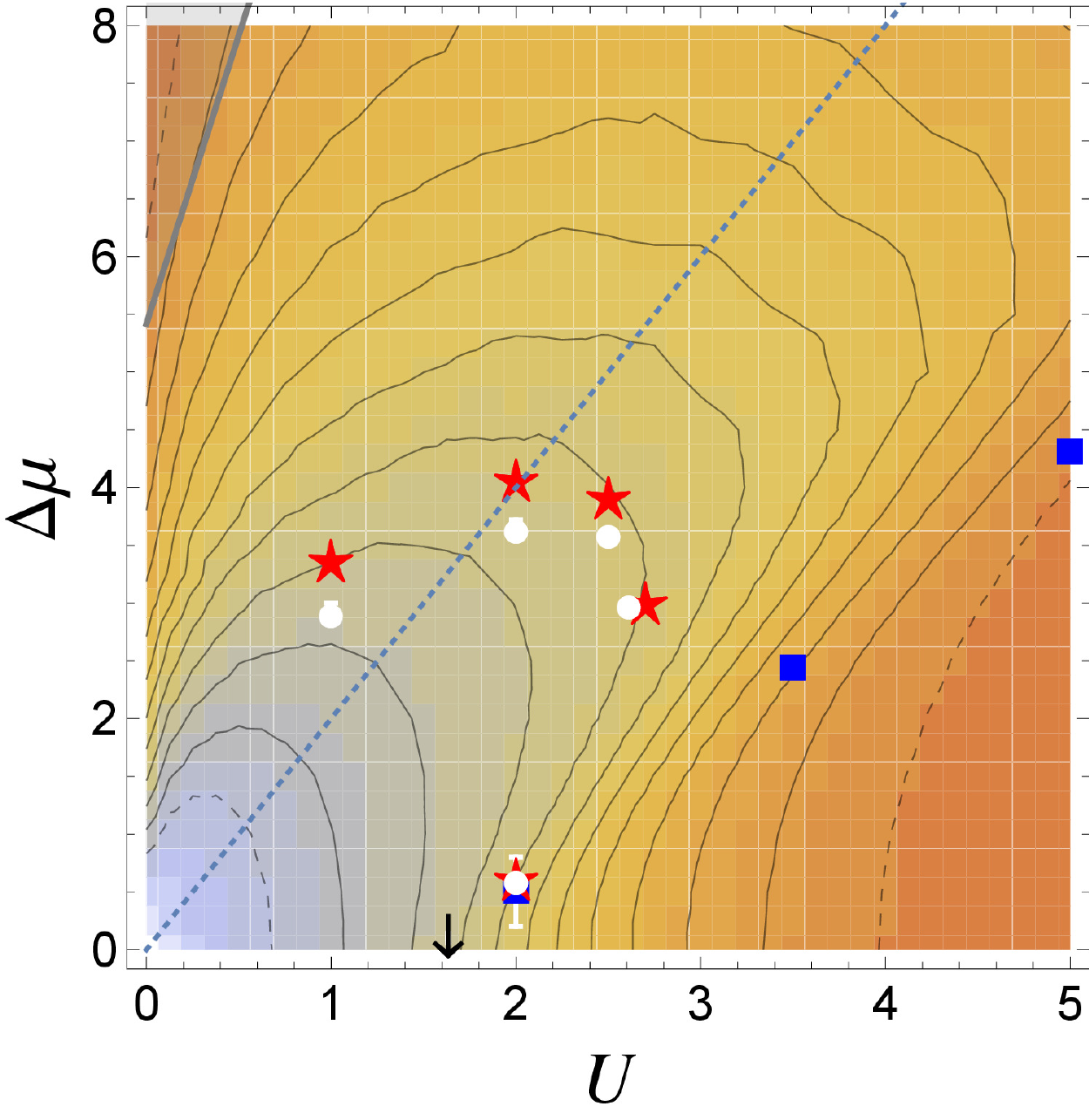}
(c)\includegraphics[width=0.22\textwidth]{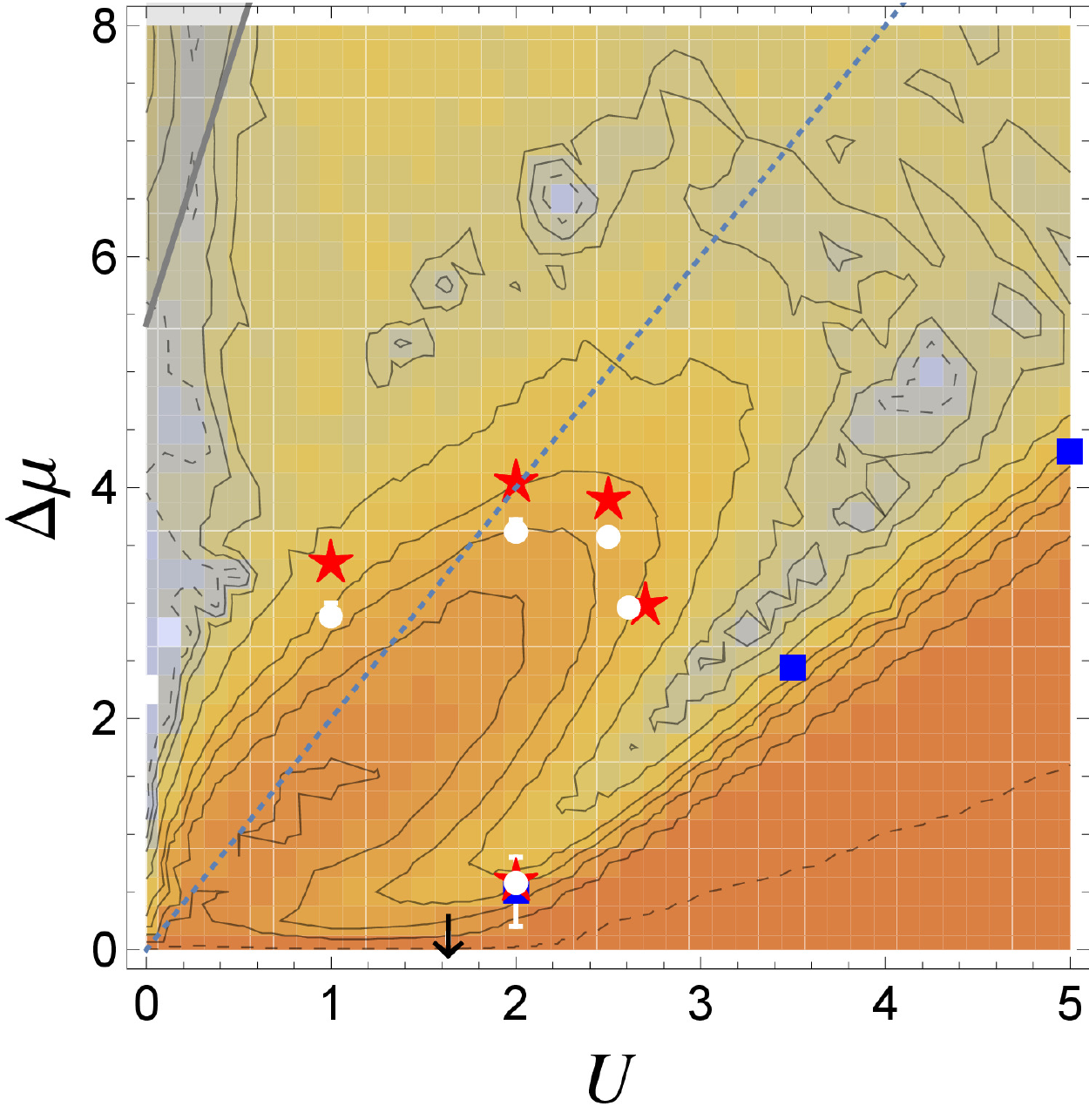}
(d)\includegraphics[width=0.22\textwidth]{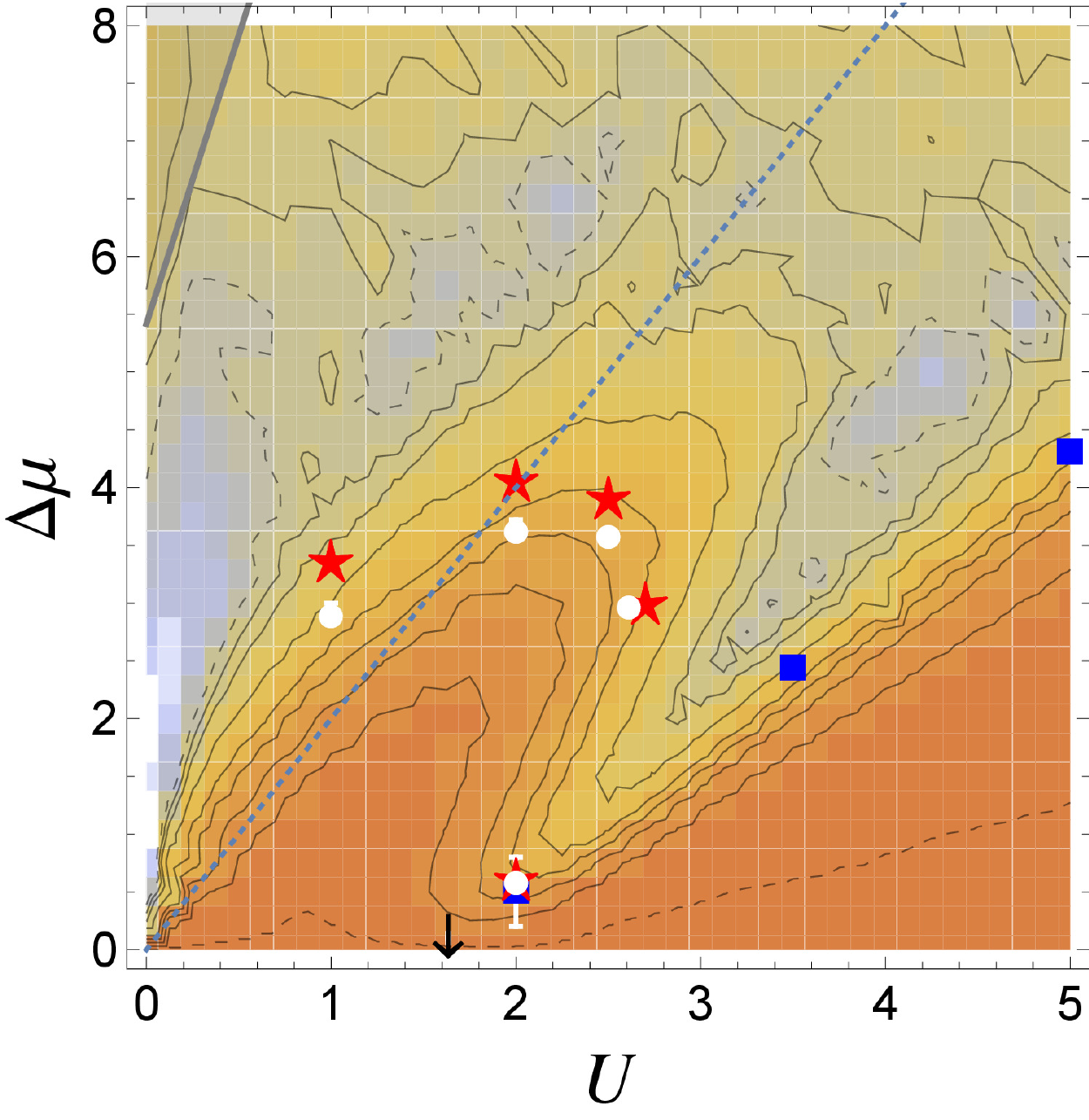}}
\caption{(Color online) 
Phase diagrams for the disordered Bose-Hubbard model at $N/L=1$ given as contour plots of (a) $\zeta$, (b) $S_{\mathrm{A}|\mathrm{B}}$, (c) $\Delta \zeta$ and (d) $\Delta S_{\mathrm{A}|\mathrm{B}}$. 
The color shading goes from low (orange/dark) to high (blue/white) value and its coarse-graining reflects the $(U,\Delta \mu)$ resolution of our calculations for $L=50$. 
The contour lines correspond to (a) $\zeta=0.1, 0.2, \ldots, 0.9$, (b) $S_{\mathrm{A}|\mathrm{B}}= 1.4, \ldots, 0.4, 0.3$, (c) $\Delta \zeta/10^{-4}= 0.1, 1, 2, \ldots, 10$  and (d) $\Delta S_{\mathrm{A}|\mathrm{B}}/10^{-4}=0.1, 1, 2, \ldots, 10$. In all cases the two extreme contours values are shown as dashed lines. Note that regions of high $\zeta$ corresponds to low $S_{A|B}$.
The circles (white) and squares (blue) denote estimations of $K$ and $E_g$ from FSS for $L\rightarrow\infty$ while the stars (red) indicate the $K=2$ values for $L=50$ as discussed in the text. The arrow (black) denotes the expected transition in the clean case at $U_c$.
The dotted straight line indicates $\Delta\mu= 2U$.
Error bars (white) show the standard error of the mean in all cases. They are within symbol size if not shown.
We emphasize that the color shading does not directly indicate the transitions, but rather quantifies the change in entanglement measures.
The grey line highlights the start of the shaded region where the probability of $\langle n_{i} \rangle \ge 4.9$ is greater than $10^{-3}$.
%
}
\label{fig:eepds}
\end{figure*}

We show our results based on $\zeta$ and $S_{\mathrm{A}|\mathrm{B}}$ for $L=50$ in fig.\ \ref{fig:eepds}, (a) and (b), respectively. 
The superfluid, small $U\lesssim 1.5$, and the Mott insulator, $U \gtrsim 2$, are clearly distinguishable in both panels. 
The boundary of the superfluid to the Bose glass is less well defined and it is not clear that there is a Bose glass region between the Mott insulator and the superfluid --- very different wavefunctions give similar average entanglement entropy. 
Following on from our prior discussion of fig.\ \ref{fig:eespecs},  we also plot in fig.\ \ref{fig:eepds}(c) the standard error\footnote{We note that for larger system sizes the variance or standard deviation may be better measures of distribution width as they do not approach zero in the infinite system limit.} of $\zeta$, $\Delta \zeta$, and, similarly, (d) $\Delta S_{\mathrm{A}|\mathrm{B}}$.
In these plots the phases become clear and their boundaries are consistent with earlier work \cite{RapSZ99}.
In particular, a Bose glass phase can be easily identified between Mott insulator and superfluid. 
Furthermore, we see that the contours for $\zeta$ and $S_{\mathrm{A}|\mathrm{B}}$ in fig.\ \ref{fig:eepds} are qualitatively similar, just as those for $\Delta \zeta$ and $\Delta S_{\mathrm{A}|\mathrm{B}}$.
We emphasize that for the entanglement-based measures present here, it is in fact 
possible to discern all of the phases with just a \emph{single} DMRG run for each $(U, \Delta \mu$, disorder realisation) data point. This is a clear advantage in terms of numerical costs when compared to calculations based on $E_g$, $\rho_s$ or $K$. 

In order to augment the finite-size phases identified in fig.\ \ref{fig:eepds}, we now perform runs with larger $L$ and employ FSS. 
To find the superfluid-Bose glass transition in the thermodynamic limit we calculate $K$ for various points along the boundary for system sizes $L = 30$, $50$, $100$ and $150$. The transition is of KT type at $K=3/2$. The corresponding points in $(U, \Delta\mu)$ which are shown as filled circles in fig.\ \ref{fig:eepds}.
For reference, we also plot the points where $K=3/2$ for $L=50$ (stars).
Similarly, the superfluid-Mott insulator transition point $U_c$ is the point on the zero disorder axis where $K=2$. We estimate it value as $U_{c} = 1.634 \pm 0.002$. 
We also calculate $E_g$ for the same system sizes and use FSS to find the Mott insulator-Bose glass boundary indicated as squares in fig.\ \ref{fig:eepds}.
The superfluid region we find is significantly smaller than that of ref.\ \cite{RapSZ99} but matches ref.\ \cite{ProS98}; the position of Mott insulator-Bose glass boundary is very similar to \cite{RapSZ99} and different to \cite{ProS98}.

The RG analysis of Refs.\ \cite{GiaS87} and \cite{RisPLG12} suggests that there may be a further \emph{Anderson glass} phase in the low $U < \Delta\mu / 2$ region of the Bose glass phase highlighted by the dashed line in fig.\ \ref{fig:eepds}. 
This would imply a critical point along the superfluid boundary at which point $K$ at the transition becomes disorder dependent.
Our entanglement analysis shows no sign of such a transition either within the Bose glass phase or on the boundary with the superfluid.
However, when $U \ll \Delta\mu$ the truncation of the basis, i.e. $\max (n_i) \leq 5$, becomes more problematic so we cannot rule out the existence of another phase in this region. 

\subsection{Density = 1/2}

\begin{figure*}[bt]
\center{
(a)\includegraphics[width=0.22\textwidth]{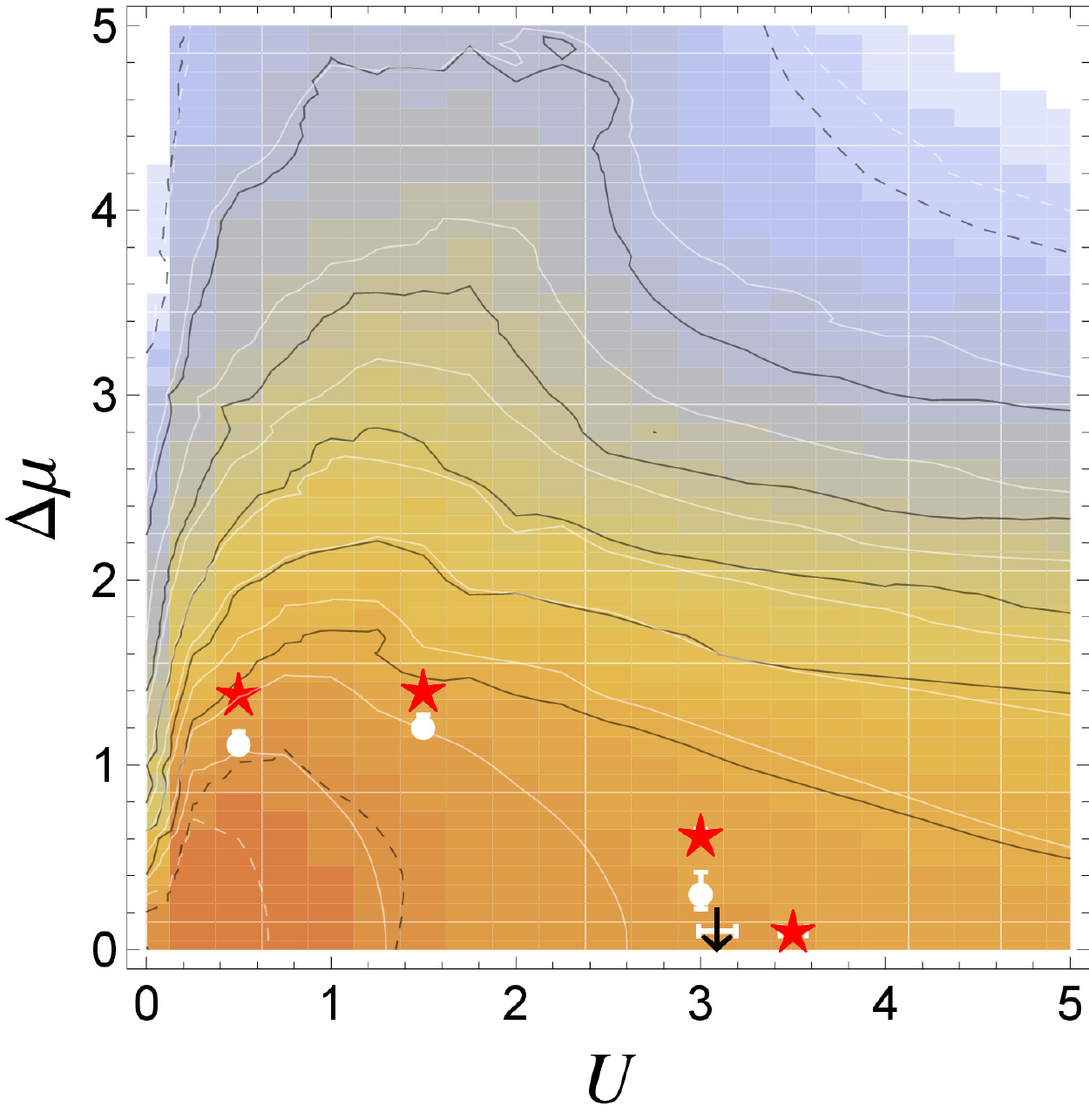}
(b)\includegraphics[width=0.22\textwidth]{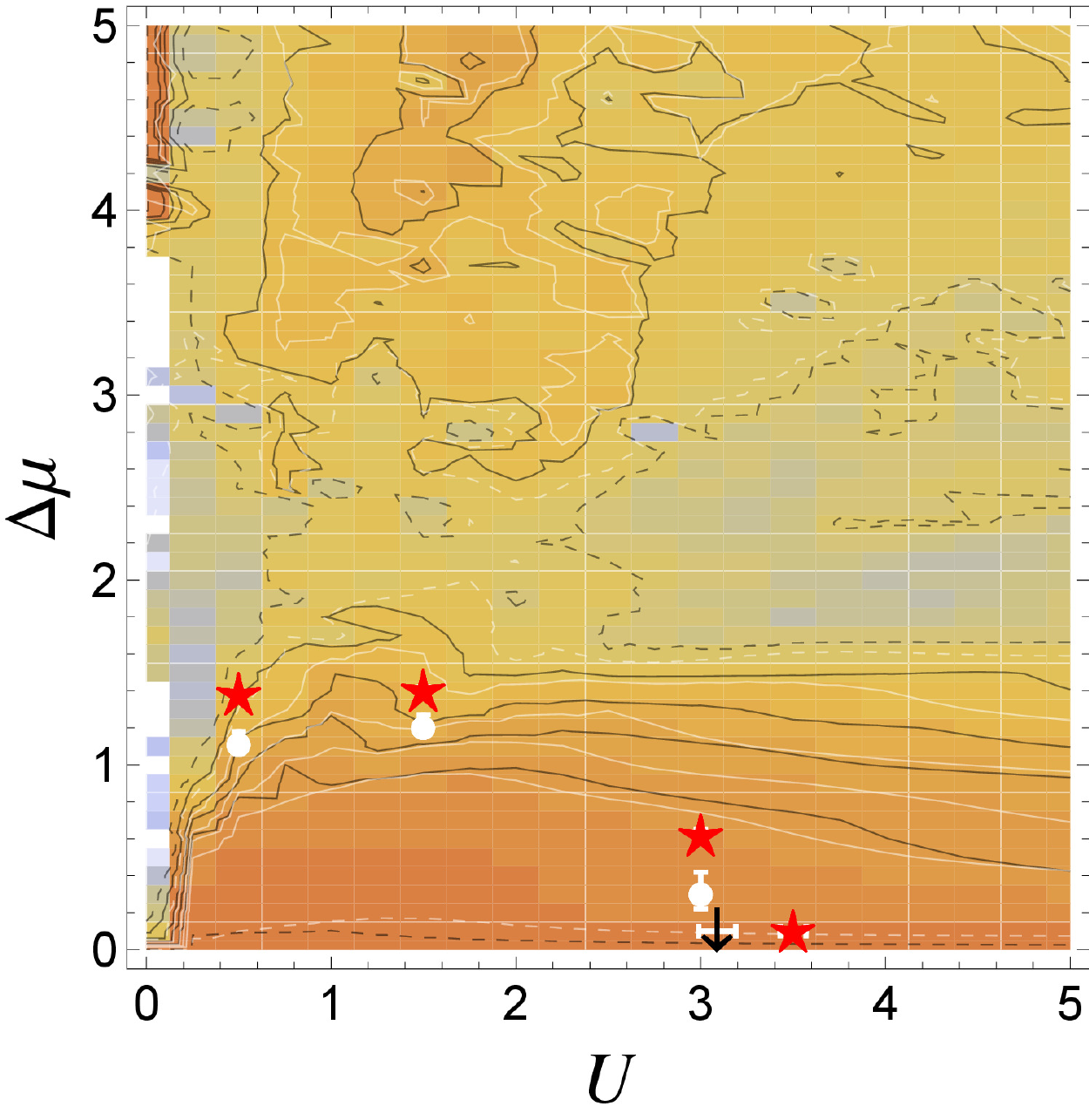}
(c)\includegraphics[width=0.22\textwidth]{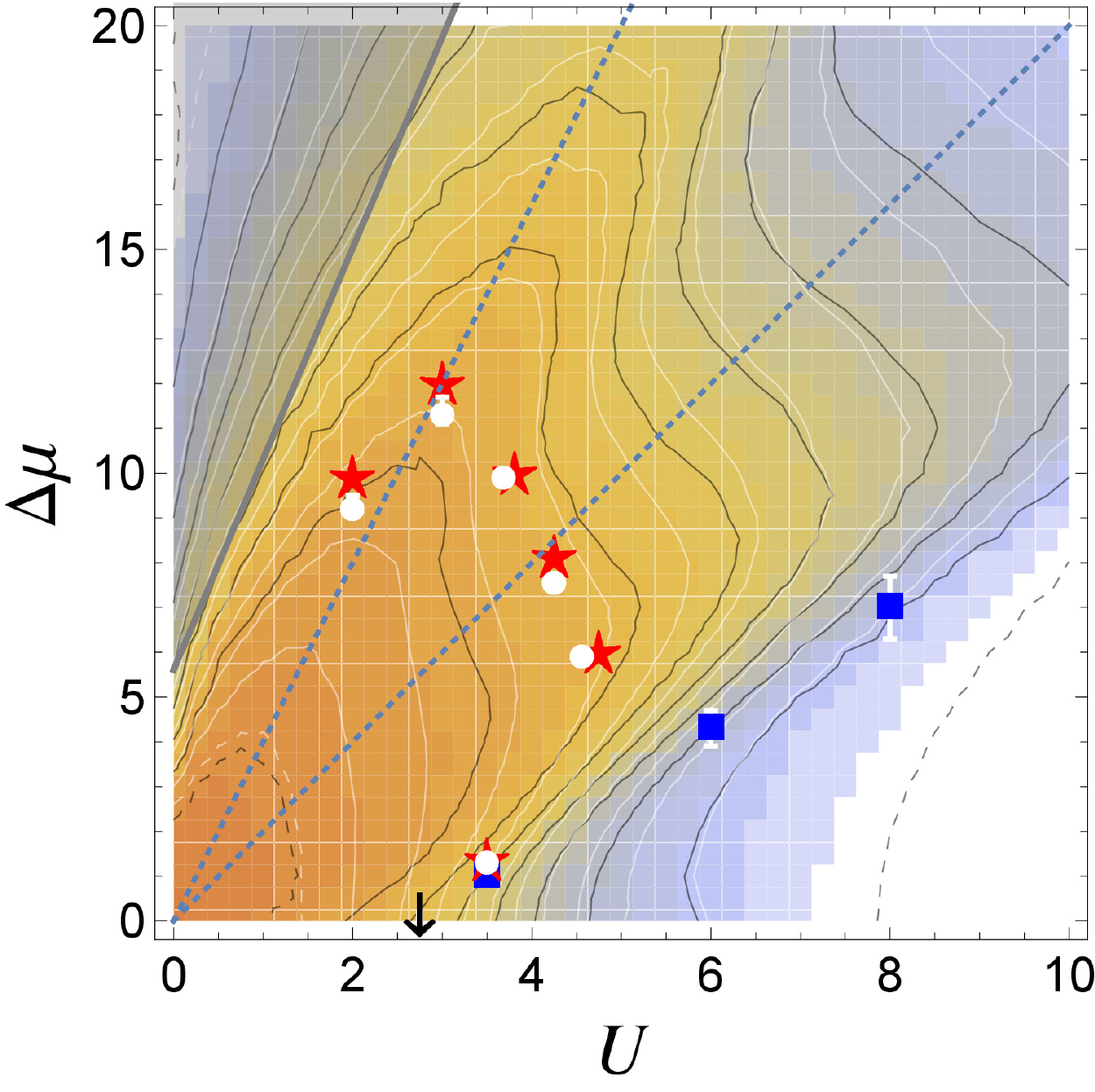}
(d)\includegraphics[width=0.22\textwidth]{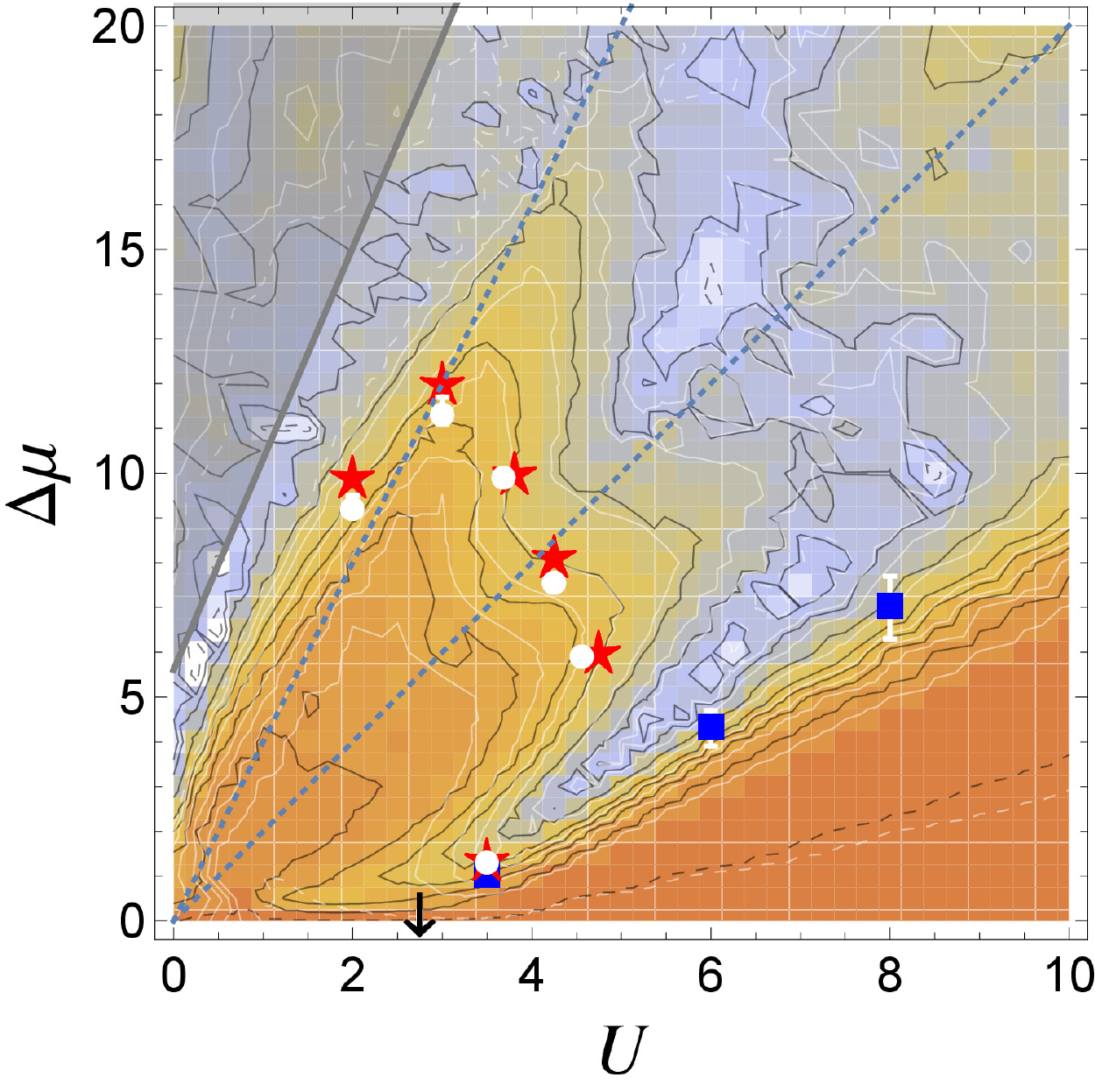}
}
\caption{(Color online) 
Phase diagrams for the disordered Bose-Hubbard model at (a+b) $N/L=1/2$ and (c+d) $N/L=2$ given as contour plots of (a+c) $\zeta$ and $S_{A|B}$, (b+d) $\Delta \zeta$ and $\Delta S_{A|B}$. 
Colors, symbols and lines (solid and dashed) denote corresponding estimates as in fig.\ \ref{fig:eepds}.  
The black contour lines correspond to (a) $\zeta=0.2, 0.3, \ldots, 0.8$, (b) $\Delta \zeta/10^{-3}= 0.1, 1, 1.5, 2, 3, 4$ for $N/L=1/2$ and (c) $\zeta= 0.1, 0.2, \ldots, 0.9$, (d) $\Delta \zeta/10^{-4}= 0.1, 1, 2, 3, 4, 6, 8, 9, 10$ for $N/L=2$.
The white contour lines represent the results for (a+c) $S_{A|B}$, (b+d) $\Delta S_{A|B}$. 
For $S_{A|B}$, the contour values are $1.2, \ldots, 0.4, 0.3$ in (a), while they are $1.4, \ldots, 0.3, 0.2$ for (c);
for $\Delta S_{A|B}$ the values are as for $\Delta\zeta$. High $\zeta$ corresponds to low $S_{A|B}$. 
The black arrow corresponds to $U_c$ for both densities as discussed in the text.
The two dotted straight lines indicates $\Delta\mu= 2U$ and $4U$.
The grey line and area have the same meaning as in fig.\ \ref{fig:eepds}.
%
}
\label{fig:zeta_eepds_half_two}
\label{fig:eepds_half}
\label{fig:eepds_two}
\end{figure*}

\begin{figure}[bt]
\center{
(a)\hspace*{-2ex}\includegraphics[width=0.22\textwidth]{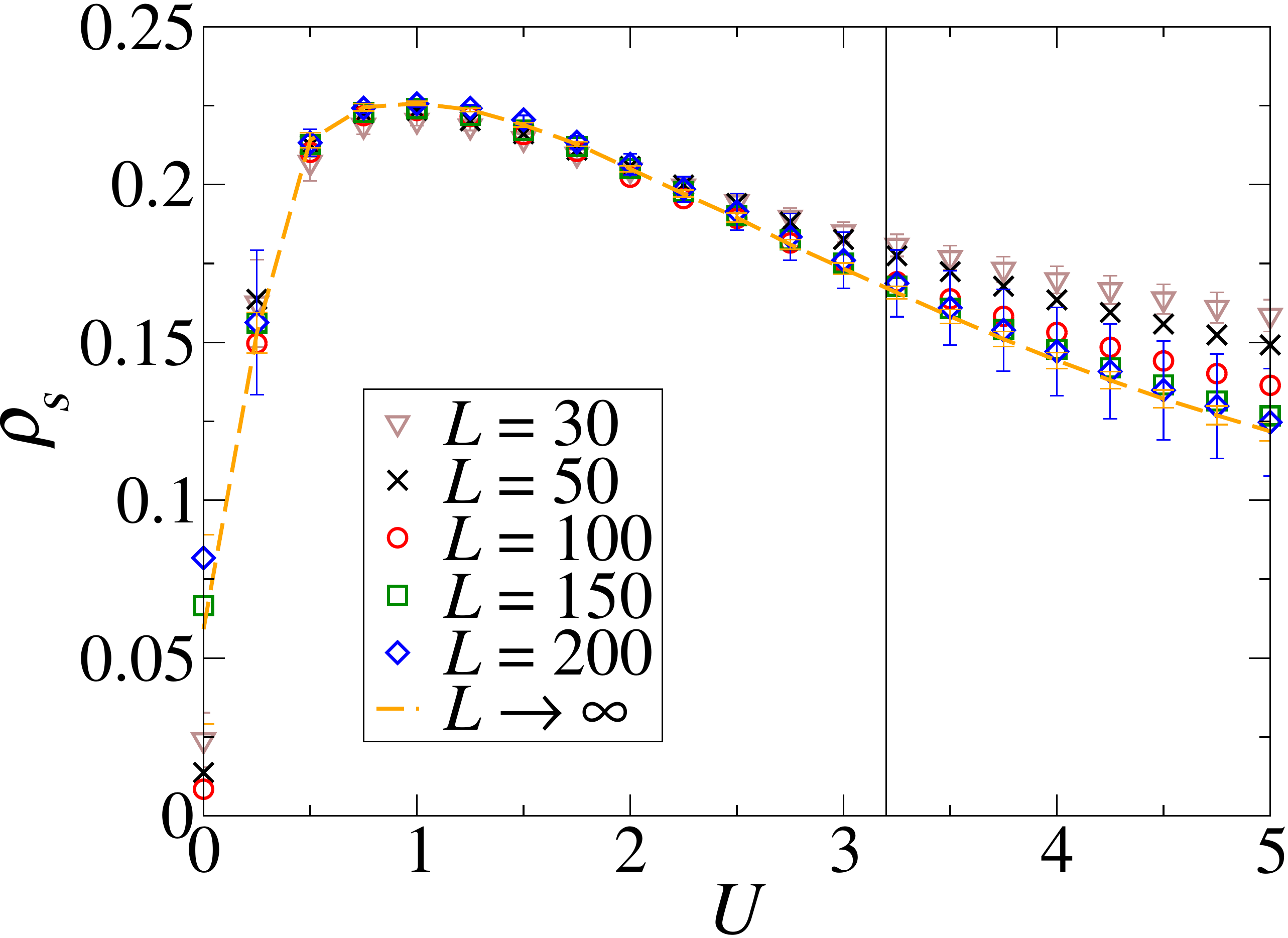}}
(b)\hspace*{-2ex}\includegraphics[width=0.22\textwidth]{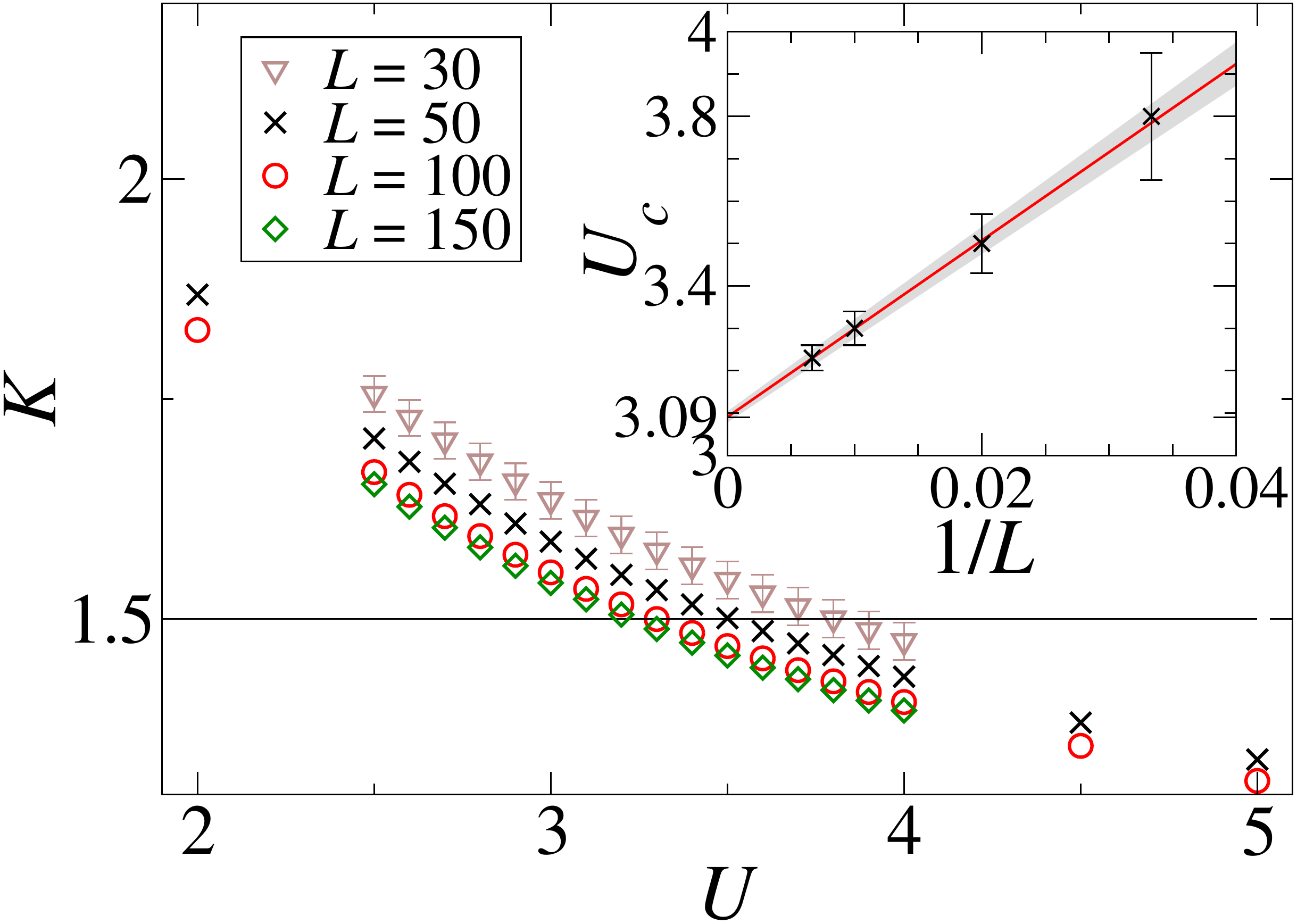}
\caption{(Color online) 
(a) Superfluid fraction $\rho_{s}(U)$ for $N/L = 1/2$ with $\Delta\mu = 0.5$ for lengths $50$--$200$. The vertical line indicates $U_c= 3.09$. The inverted triangles give the finite-size scaled $L\rightarrow\infty$ limit with the dashed line a guide to the eye.
(b) Luttinger parameter $K$ for various lengths $30$--$150$ at $N/L=1/2$. The horizontal line highlights $K = 3/2$. The inset shows the FSS analysis. 
}
\label{fig:stiffness_half}
\end{figure}

The clean case for $N/L=1/2$ remains a superfluid for all values of $U$ \cite{FisWGF89}.
When $\Delta\mu$ is increased, our entanglement measures indicate the eventual emergence of a Bose glass phase as shown in fig.\ \ref{fig:eepds_half}(a+b). 
%
Still, the superfluid phase for $L=50$ seems to extend up to $\Delta\mu \lesssim 1$ for $U \lesssim 5$ as shown by all four entanglement measures.
The Giamarchi-Schulz criterion \cite{GiaS87,RapSZ99} implies that the Bose-Hubbard model should be in a Bose glass phase for $K<3/2$. In fig.\ \ref{fig:eepds_half}(a+b) we show that the resulting boundaries indicate that the superfluid phase extends as far as $U_{K=3/2}=3.5\pm 0.1$, i.e.\ it ends somewhat earlier for low $\Delta\mu$ than suggested by our entanglement measures.
In order to explore this region further, we have also calculated $\rho_{s}$ for fixed $\Delta\mu = 0.5$ and sizes $L=50$, $100$, $150$, and $200$ as shown in fig.\ \ref{fig:stiffness_half}(a).
The results for $\rho_{s}$ have been computed for increased bond dimension $\chi=400$ with $40$ DMRG sweeps and $20$ disorder configuration to offset the reduction in precision due to periodic boundaries.
The figure shows that for $U\gtrsim 3$, $\rho_{s}$ decreases when increasing $L$ as expected in the Bose glass phase. However, the decrease is very slow and, for the system sizes attainable by us, even seems to saturate at non-zero values. 
These results suggest that for finite systems, the $K=3/2$ criterion significantly underestimates the extent of the superfluid phase, while our four entanglement measures and $\rho_{s}$ predict a much larger region.
Performing a FSS analysis for $K$ as shown in fig.\ \ref{fig:stiffness_half}(a) we find the $U$ values, for which $K = 3/2$ in the limit $L\rightarrow\infty$, converge towards a limiting value of $U_{c} = 3.09 \pm 0.01$ (see also fig.\ \ref{fig:eepds_half}). This again indicates that in an infinite system, we expect the superfluid to Bose glass transition to take place at much lower values of $U_c$ than observed for $L=50$. 
The relevance of this result is of course that experimental realizations of the Bose-Hubbard model are typically in cold atom systems, which are limited to finite system sizes, currently a typical lattice dimension is $\sim 50-100$ \cite{GreMEH02}.

For values of $\Delta\mu\gtrsim 1$, the situation is less severe and we see in fig.\ \ref{fig:eepds_half}(a+b) that our entanglement-based measures again qualitatively agree with the Giamarchi-Schulz criterion, both for $L=50$ and estimated via FSS at $L\rightarrow\infty$.

\subsection{Density = 2}
To the best of our knowledge, the phases for $N/L = 2$ have not been shown before in the literature. 
Due to our numerical restriction of five bosons per site, this regime is close to the limit of what can be studied reliably, particularly for small $U$ where the occupancy per site should be large. 
For large $U$, one might expect that we will have a Mott insulator of boson \emph{pairs}, while a superfluid of boson pairs emerges for small $U$ and small $\Delta\mu$. 
As before, we envisage a disordered Bose glass phase for large $\Delta\mu$. 
With more particles per site than in the $N/L=1$ case, we could furthermore expect that onset of the Mott transition at $\Delta\mu=0$ is at larger values of $U$, since there is a larger energy penalty to pay for a doubly occupied site \cite{FisWGF89}. 
Similarly, as the cost for two boson pairs to go onto the same site is $2U$, we expect the $2 U = \Delta\mu/2$ line to characterize the superfluid phase as in the $N/L=1$ case. In addition, one might conjecture to see a remnant of the $U = \Delta\mu/2$ condition.

In fig.\ \ref{fig:eepds_two}(c+d), we show that our expectations are largely validated. In particular, a double lobe shape for the superfluid phase emerges and allows a possible re-entrant behaviour given a suitable cut across parameter space. The gradient of the Mott insulating phase boundary is shallower ($\sim 4/3$) when compared to $N/L=1$. 
Furthermore, both the $\zeta$ and $S_{\mathrm{A}|\mathrm{B}}$ based entanglement measures, as well as their errors, $\Delta\zeta$ and $\Delta S_{\mathrm{A}|\mathrm{B}}$, capture the phases equally well and agree with the $K$ and $E_g$ estimates. 
Note that for $N/L=2$, the KT superfluid-to-Mott transition at $\Delta\mu=0$ corresponds to $K=2$ and we finite-size scale the Luttinger parameter to find $U_{c} = 2.75 \pm 0.03$.

We emphasize that the points for small $U$, see top left of the phase diagram in fig.\ \ref{fig:eepds_two}(c+d), should be viewed with caution as the basis truncation will affect the results.
The grey line in fig.\ \ref{fig:eepds_two} --- as in fig.\ \ref{fig:eepds} --- indicates the points at which the probability of obtaining a site with $\langle n_{i} \rangle \ge 4.9$ reaches $10^{-3}$.
This clearly shows that for the Bose glass with small $U$ all wavefunctions are beginning to reach the limit five of bosons per site, however in the bulk of the phase diagram the results are not affected.


\section{Conclusion}
We have analysed the phase diagrams of the disordered Bose-Hubbard model for fillings $N/L = 1/2$, $1$ and $2$ using the entanglement-based measures $\zeta$, $S_{\mathrm{A}|\mathrm{B}}$, $\Delta\zeta$ and $\Delta S_{\mathrm{A}|\mathrm{B}}$. 
We find that despite success in ref.\ \cite{DenCOM13}, $\zeta$ or $S_{\mathrm{A}|\mathrm{B}}$ alone do not always faithfully reproduce the phase diagrams.
The error-based measures, $\Delta\zeta$ and $\Delta S_{\mathrm{A}|\mathrm{B}}$, provide a much clearer picture --- the distributions of the values contain more information regarding the nature of the phase than the mean values alone.
These measures are an excellent means of quickly identifying the different phases of the system while removing the need for multiple DMRG runs per measurement and special boundary conditions.
Unfortunately, they do not seem to exhibit a simple FSS behavior, at least for the system size up to $L=200$ used here. While $\zeta$ and $S_{\mathrm{A}|\mathrm{B}}$ and, in particular, $\Delta\zeta$ and $\Delta S_{\mathrm{A}|\mathrm{B}}$ provide a numerically convenient, qualitative outline of the phase boundaries, it seems still necessary to apply FSS to $K$ and $E_g$ for estimates of the boundaries in the $L\rightarrow\infty$ limit.
For $N/L = 1$ our phase diagram is found to complement the results of refs.\ \cite{ProS98,RapSZ99}.
For $N/L = 1/2$ the diagram shows strong finite-size effects and the critical $U$ defined by the Giamarchi-Schulz criterion is not apparent for these finite systems. 
Finally, for $N/L = 2$ the superfluid phase has a \emph{double-lobed} appearance giving rise to re-entrance phenomena. 

\acknowledgments
We would like to thank Miles Stoudenmire for help with {\sc ITensor}. We gratefully acknowledge discussions with Kai Bongs and Nadine Meyer. We are grateful to the EPSRC for financial support (EP/J003476/1) and provision of computing resources through the MidPlus Regional HPC Centre (EP/K000128/1).

The supporting data for this research is openly available from the University of Warwick research archive portal at http://wrap.warwick.ac.uk/71189/.

\end{document}